\documentstyle[12pt,aps,floats,epsf]{revtex}
\newcommand{\la}{\langle}
\newcommand{\ra}{\rangle}
\newcommand{\be}{\begin{equation}}
\newcommand{\ee}{\end{equation}}
\renewcommand{\baselinestretch}{1.0}
\begin{document}
\title{{\small April 3,\ 1995 \hspace*{100mm}TPR-95-04}\\[6mm]
\LARGE\bf Rapidity Dependence of Strange\\Particle Ratios\\in
 Nuclear Collisions\\[6mm]}
\author{Claus Slotta$^a$, Josef Sollfrank$^b$ and Ulrich Heinz$^a$\thanks{Work
supported by GSI, DFG and BMFT.}\\[4mm]}
\address{$^a$Institut f\"ur Theoretische Physik, Universit\"at Regensburg\\
D-93040 Regensburg, Germany}
\address{$^b$Research Institute for Theoretical Physics\\
P.O.Box 9 (Siltavuorenpenger 20 C)\\
FIN-00014 University of Helsinki, Finland}
\maketitle
\vspace*{10mm}
\begin{abstract}

\centerline{\bf ABSTRACT}\vspace*{4mm}
It was recently found [1,2] that in sulphur-induced nuclear collisions
at 200 A GeV the observed strange hadron abundances can be explained
within a thermodynamic model where baryons and mesons separately are
in a state of relative chemical equilibrium, with overall strangeness
being slightly undersaturated, but distributed among the strange
hadron channels according to relative chemical equilibrium with a
vanishing strange quark chemical potential. These studies were made
either in a narrow central rapidity window [1] or (in the case of S-S
collisions [2]) also using the rapidity-integrated total multiplicities.

We develop a consistent thermodynamic formulation of the concept of
relative chemical equilibrium and show how to introduce into the
partition function deviations from absolute chemical equilibrium, e.~g.~an
undersaturation of overall strangeness or the breaking of chemical equilibrium
between mesons and baryons.
We then proceed to test on the available data the hypothesis that the strange
quark chemical potential vanishes everywhere, and that the rapidity
distributions of all the observed hadrons can be explained in terms of one
common, rapidity-dependent function $\mu_{\rm q}(\eta)$ for the baryon
chemical potential only. The aim of this study is to shed light on the
observed strong rapidity dependence of the strange baryon ratios in
the NA36 experiment [3].
\end{abstract} \newpage

\section*{Introduction}
In the meantime it is almost commonly agreed that fireballs resulting from
collisions between heavy ions probably live long enough to establish a state of
local thermal and at least relative [1,2] chemical equilibrium. But according
to
our knowledge no thermodynamic description has been developed yet which gives a
quantitative explanation for the different shapes of the rapidity
distributions.

We introduce a simple hydrodynamical model of fireballs resulting from S+S
collisions at 200$\,$A$\,$GeV and allow as a new feature a dependence of the
chemical potential on the space-time rapidity. We show that this simple
extension increases dramatically the agreement with experimental spectra.

We use the complete set of rapidity distributions for $\pi^-$, K$^\pm$,
K$^0_s$,
$\Lambda$, $\bar{\Lambda}$ and p-$\bar{\rm p}$ from 200$\,$A$\,$GeV/c S+S
collisions obtained by the NA35 collaboration [4]. Where applicable
we back-extrapolate the published data to the experimental acceptance windows
(by inverting the algorithm given by NA35 [4]) since our model is able to take
them into account explicitly. In the figures below the index ``e'' is short for
``within experimental acceptance''.
\section*{Relative Chemical Equilibrium}
J.~Rafelski et al.~[1] suggested a simple method to test experimental data for
possible signals of a QGP in the early stages of heavy ion collisions based on
a
chemical analysis of the final hadronic composition. The underlying concept
rests on the assumption of local thermal and absolute chemical equilibrium with
respect to up- and down-quarks, at least in a region near central rapidity.
Absolute chemical equilibrium of { \it strange} particles is not expected due
to
the rather high mass of $s\bar{s}$-pairs. Instead one assumes the weaker
condition of relative chemical equilibrium: The strange phase space is not
fully
saturated, but strangeness has been distributed among the available strange
hadron channels according to equilibrium fugacities, maximizing the entropy. A
convenient way of parametrizing this is by introducing a common saturation
factor $0<\gamma_{\rm s}\le 1$ for both strange quarks and anti-quarks. The
abundance of hadrons is then regulated by fugacities $\lambda_k$ given by a
product of valence quark fugacities multiplied by the factor $\gamma_{\rm
s}^{|s_k|}$, where $|s_k|$ counts the number of strange quarks plus anti-quarks
in the respective hadron species.

In [1] the factor $\gamma_{\rm s}$ was introduced heuristically; here we
present
a rigorous thermodynamic formulation of the concept of relative chemical
equilibrium.

In a statistical description of an ideal gas mixture containing $M$ different
components, the one particle distribution functions $f_k\equiv f_k(x,p),\,
k=1\ldots M$ contain all of the system's information.
Given sufficient time, this distribution function will evolve to the local
equilibrium solution, whose form can be very efficiently derived by maximizing
the entropy subject to constraints from conservation laws. For thermal and
absolute chemical equilibrium the only constraints are due to the absolutely
conserved quantities energy, baryon number and (on the time scale of nuclear
reactions) strangeness. (Charge conservation in practice doesn't play a role in
nuclear collisions.) However, if different chemical processes occur at
different, well separated time scales, a state of relative chemical equilibrium
may arise at intermediate time scales where the fast processes have already
equilibrated but the slow ones are still far off equilibrium. This intermediate
state can be characterized by additional constraints on the chemical
composition
which, when implemented into the procedure of entropy maximization, yield the
local equilibrium distribution for {\it relative} chemical equilibrium.

Let us consider a small fluid cell at freeze-out in its own rest system. Its
entropy content can be written as
\be \label{entropie}
  {\cal S} = -\sum_k \int d\omega
  \Big\{ f_k\,\ln f_k +\theta_k\big(1-\theta_k f_k\big)\,
      \ln \big(1-\theta_k f_k\big) \Big\}
\ee
where the integral is over conventional phase space $\int d\omega =
\int_{\Delta
V}d^3x\int\frac{d^3p}{(2\pi)^3}$, and $\theta_k=+1 (-1)$ applies to baryons
(mesons). The conservation of energy, light ($q$) and strange ($s$) quark
flavours in strong interactions leads to the constraints
\be \label{rand}
\la E\ra = \int d\omega \sum_k E_k f_k\,,\;\;\;
\la Q\ra = \int d\omega \sum_k q_k f_k\,,\;\;\;
-\la S\ra =  \int d\omega \sum_k s_k f_k\,.
\ee
Here  $\la E\ra$ denotes the cell's average thermal energy, $\la Q\ra$ stands
for its average net light quark number (we neglect the small $u$-$d$ mass
difference), and $\la S\ra$ for its average amount of net strangeness. On the
right sides $q_k$ and $s_k$ count the number of light resp. strange valence
quarks in a hadron of species $k$; note that anti-quarks contribute with
negative signs.

The concept of a partially saturated strange phase space can be implemented by
the additional condition
\be \label{randneu}
  \la|S|\ra = \int d\omega \sum_k |s_k| f_k\;.
\ee
Since the amounts $\la S\ra$ and $\la|S|\ra$ of overall strangeness can be
specified independently, this allows for a description of both over- and
under-saturated strange phase space.
We can further parametrize
\be \label{randpara}
\la|S|\ra =\sum_k \xi_k\, \la|S|\ra_k^{\rm Eq}\;,
\ee
where  $\la|S|\ra_k^{\rm Eq}$ denotes the contribution of the hadron species
$k$
to overall strangeness in absolute chemical equilibrium. The factors $\xi_k$
represent the degree of population of the strange phase space in each hadronic
sector, which may differ from their equilibrium values $\xi_k=1$; the range
$0<\xi_k\le1$ corresponds to strangeness suppression. The parametrization
(\ref{randpara}) explicitly takes into account that quarks are bound into
hadrons at freeze-out and allows for a suppression in dependence on their
valence quark content.

The most probable distribution function $f_k$ can be found by extremizing the
functional ${\cal S}[f_k]$, subject to the constraints (\ref{rand},
\ref{randneu}). This can be done by means of Lagrangean multipliers $\beta$,
$\nu_q$, $\nu_s$, $\sigma$ and leads to
\be \label{verteilung}
  f_k = \frac{1}{e^{\beta E_k+\nu_q q_k+\nu_s s_k+\sigma|s_k|}+\theta_k}\;.
\ee
Inserting (\ref{verteilung}) into the entropy formula (\ref{entropie}) we
obtain
\be \label{entropieres}
  {\cal S}  =  \beta\la E\ra + \sum_k  (\nu_{\rm q} q_k+\nu_{\rm s} s_k)\la N_k
\ra + \ln \Xi + \sum_k \sigma |s_k|\la N_k\ra\;,
\ee
where we define the average particle number of species $k$ in the considered
volume element $  \la N_k\ra = \int d\omega f_k $ and
\be \label{lnZ}
  \ln \Xi = \sum_k \theta_k \int d\omega \ln \left(1+\theta_k e^{-(\beta E_k
  +(\nu_{\rm q} q_k+\nu_{\rm s} s_k)+\sigma|s_k|)}\right)\;.
\ee
The limit $\sigma \to 0$ corresponds to neglecting the additional condition
(\ref{randneu}) in the extremization procedure. In this case we recover from
(\ref{lnZ}) the expression for the grand canonical partition function in
absolute chemical equilibrium.

It can be easily shown that $\ln\Xi$ generates all thermodynamic quantities
($\alpha_k \equiv (\nu_{\rm q} q_k+\nu_{\rm s} s_k) +\sigma |s_k|$):
\be \label{E}
 -\frac{\partial}{\partial\beta} \ln\Xi\Big|_{\alpha_k,V} = \la E\ra\,,\;\;\;
 -\frac{\partial}{\partial \alpha_k} \ln\Xi\Big|_{\beta,V} = \la
N_k\ra\,,\;\;\;
  T\frac{\partial}{\partial V} \ln\Xi\Big|_{\alpha_k,T} = p\,.
\ee
Hence it is a thermodynamic potential.
As usual we identify $\beta=\frac{1}{T}$ and $\alpha_k = -\frac{\mu_k}{T}$.
The chemical potentials in absolute chemical equilibrium
\be
 \mu_k^{\rm Eq} = -T(\nu_{\rm q} q_k + \nu_{\rm s} s_k)
\ee
are connected with $\mu_k$ via
\be \label{muk}
  \mu_k = \mu_k^{Eq} -T\sigma|s_k| \equiv \mu_k^{Eq}+\tilde{\sigma}|s_k|\;,
\ee
where we defined $\tilde{\sigma} = -T\sigma$.

In Boltzmann approximation we can determine the suppression factors $\xi_k$ by
comparing (\ref{randneu}) and (\ref{randpara}) as
$\xi_k = e^{(\beta\tilde{\sigma}|s_k|)}$. They are directly related to the
saturation factor $\gamma_{\rm s}$ via $\xi_k=\gamma_{\rm s}^{|s_k|}$:
\be \label{xi}
0 < \gamma_{\rm s}^{|s_k|} = \xi_k = (e^{\beta\tilde{\sigma}})^{|s_k|} \le 1\;.
\ee
An overall suppression of the strange quark phase space has no influence on the
thermodynamic relations (\ref{E}); it only enters into the relation
(\ref{entropie}) for the entropy.
If decomposed into particle specific contributions ${\cal S}=\sum_k {\cal S}_k$
we immediately see that each component receives an additional contribution
$-|s_k|\ln\gamma_{\rm s}\la N_k\ra$.

The starting point of our formalism were the suppression factors $\xi_k$
weighting each hadron species separately. As a consequence of the entropy
maximization, however, we automatically arrived at the concept of relative
chemical equilibrium with {\rm one common} factor $\gamma_{\rm s}$. This
justifies a posteriori the parametrization of Rafelski et al.~[1].

The procedure is easily generalized to allow for a meson-baryon non-equilibrium
as suggested in the last of Refs.~[1]. This might be necessary if processes of
the type $M$+$M$ $\leftrightarrow$ $\bar{B}$+$B$ are slow. Let us assume that
baryons are in absolute chemical equilibrium and mesons are not. This requires
the inclusion of the additional constraint
\be \label{randneuneu}
\la M\ra = \sum_k\,\mbox{$\frac{1}{2}$}(1-\theta_k)\,\int\,d\omega\,f_k\;,
\ee
where again we can parametrize in the same spirit as (\ref{randpara})
\be
\la M \ra =\sum_k\,\mbox{$\frac{1}{2}$}(1-\theta_k)\,\xi^{\rm m}_k\,\la N_k
\ra^{\rm Eq}\;.
\ee
Combining (\ref{randneu}) and (\ref{randneuneu}) we find the relative
equilibrium distribution
\be
f_k = \gamma_{\rm s}^{|s_k|}\,\gamma_{\rm
m}^{\frac{1}{2}(1-\theta_k)}\,f_k^{\rm
Eq}\;,
\ee
where in Boltzmann approximation $\xi_k^{\rm m}$ and $\gamma_{\rm m}$ are
related by $\xi_k^{\rm m} = \gamma_{\rm m}^{\frac{1}{2}(1-\theta_k)}$. The case
where also the baryons are suppressed or enhanced relative to their equilibrium
abundance can be treated in an analogous way.
\section*{Calculating the Spectra}
For the momentum spectra of hadrons emitted from the fireball at freeze-out we
use the simple hydrodynamical model developed by E.~Schnedermann et al.~[5]. It
is based on the idea of a thermal distribution with one freeze-out temperature
$T_{\rm f}$ characterizing the instant at which fluid cells loose their thermal
contact to the fireball. Experiments find similar exponential slopes in all
$m_\bot$-spectra, independent of the rapidity window [4]. This supports the
interpretation of freeze-out at one common and constant temperature.

The general expression for the invariant momentum distribution is [6]
\be \label{cooper}
E\frac{d^3N}{dp^3} = \frac{g}{(2\pi)^3} \int_{\sigma_{\rm f}} f(x,p)\,p^\nu
d^3\sigma_\nu\;,
\ee
where $g$ is the spin degeneracy of the hadrons under consideration and
$\sigma_{\rm f}$
is the 3-dimensional freeze-out hypersurface in space-time, along which the
condition $T = T_{\rm f}$ is fulfilled.  We choose a cylindrical coordinate
system  $(r,\phi,z)$
which best reflects the global symmetries of the expanding fireball and obtain
for the normal vector of the freeze-out surface
\be 
d^3\sigma_\nu =
\left( 1,-\frac{\partial t_{\rm f}}{\partial r} \cos\phi ,
-\frac{\partial t_{\rm f}}{\partial r} \sin\phi , -\frac{\partial t_{\rm f}}
{\partial z} \right)\,rdr\,d\phi\,dz\;.
\ee
In an azimuthally symmetric geometry of this kind it is practical to decompose
the velocity field in the following way [7]:
\be \label{umue}
u^\nu = \left( \cosh\rho\,\cosh\eta,\hat{e}_\bot \sinh\rho,\cosh\rho\,
\sinh\eta\right)\;.
\ee
Here $\eta=\eta(t,r,z)$ is the longitudinal flow rapidity by which each volume
element on the z-axis moves relative to the center of mass, and
$\rho=\rho(t,r,z)$ is the rapidity corresponding to the transverse flow of the
volume element at position $(r,z)$ as seen from a reference point at $z$ on the
beam axis moving with the local flow velocity there. The 4-momentum can be
parametrized in terms of rapidity $y$ and transverse mass $m_\bot =
\sqrt{p_\bot^2+m_0^2}$ as
\be 
p^\nu = (m_\bot \cosh y,p_\bot \cos\phi_{\rm p},p_\bot \sin\phi_{\rm p},m_\bot
\sinh y)\;.
\ee
After suitably orienting the coordinate system we obtain in
Boltzmann-approximation:
\begin{eqnarray} 
\frac{d^2N}{\,dy\,dm_\bot^2} & = & \frac{g}{4\pi}\,m_\bot \int_{t_{\rm f}}
dz\,r
\,dr\,e^{\mu/T}\,
e^{-\tilde{\alpha}\cosh(\eta-y)} \nonumber \\
& & \times \left\{ \left( \cosh y -\sinh y\frac{\partial t_{\rm f}}{\partial z}
\right)
{\rm I}_0(\alpha) - \frac{p_\bot}{m_\bot}\frac{\partial t_{\rm f}}{\partial r}
\,{\rm I}_1(\alpha) \right\}\;,
\end{eqnarray}
where $\tilde{\alpha} = (m_\bot/T)\cosh\rho$ and
$\alpha = (p_\bot/T)\sinh\rho$.
The 2-dimensional integration extends along the curve determined by all points
$(r,z)$ which at time $t_{\rm f}(r,z)$ satisfy $T(t_{\rm f}(r,z),r,z) =
T_{\rm f}$.

For further evaluation we introduce new variables and transform $(t,r,z)$ into
$(\tau,r,Y)$ with the longitudinal proper time
$\tau = \sqrt{t^2 - z^2}$
and the space-time rapidity
$Y= 0.5 \ln \left[(t+z)/(t-z)\right]$.
Numerical simulations of the space-time evolution of the hot zone in
ultrarelativistic nuclear collisions show an almost linear velocity profile in
beam direction [8]. Hence a boost invariant scenario, if restricted to the
fireball's finite extension, is a good approximation for the longitudinal fluid
dynamics and leads to $Y\equiv\eta$, i.e.~the identity of space-time rapidity
$Y$ and fluid rapidity $\eta$.
The partial derivatives $\frac{\partial t_{\rm f}}{\partial r}$ and
$\frac{\partial t_{\rm f}}{\partial z}$ are easily evaluated if we parametrize
the freeze-out hypersurface by a proper time surface $\tau_{\rm f}(r)$:
$t_{\rm f}(r,z) = \sqrt{\tau_{\rm f}^2(r)+z^2}$.
In doing so we simultaneously express the invariance of the decoupling process
against longitudinal boosts and obtain for the rapidity distribution  using
$z=\tau_{\rm f}(r)\sinh\eta$:
\begin{eqnarray} \label{result}
\frac{dN}{dy} & = & \frac{g}{4\pi}\,\int_{m_\bot^{\rm lo}}^{m_\bot^{\rm hi}}\,
dm_\bot^2\,\int_{-\tilde{\eta}}^ {\tilde{\eta}}\,d\eta \int_0^{R_{\rm f}}\,dr\,
\tau_{\rm f}(r)\,r\,e^{\mu/T}e^{-\tilde{\alpha}\cosh(y-\eta)}\\ \nonumber
& \times & \left(m_\bot\cosh(y-\eta)\,{\rm I}_0(\alpha) - p_\bot\frac{\partial
\tau_{\rm f}(r)}{\partial r}\,{\rm I}_1(\alpha) \right)\;.
\end{eqnarray}
In this expression $m_\bot^{\rm lo},\;m_\bot^{\rm hi}$ denote the experimental
limits in which the spectrum was measured,
the freeze-out radius is called $R_{\rm f}$ and the longitudinal extent of the
fireball is fixed via the finite interval $-\tilde{\eta}\ldots\tilde{\eta}$.
We will later use the approximation where we assume transversally instantaneous
freeze-out ($\frac{\partial\tau_{\rm f}(r)}{\partial r}=0$) and replace the
transverse flow velocity profile by its radial average. In this case
eq.~(\ref{result}) reduces to the simpler form
\begin{eqnarray} \label{endresult}
\frac{dN}{dy} = \frac{g\,\tau_{\rm f}R_{\rm f}^2 }{8\,\pi}\,\int_{m_\bot^{\rm
lo}}^{m_\bot^{\rm hi}}\,dm_\bot^2\, m_\bot\,{\rm I}_0(\alpha)
\int_{-\tilde{\eta}}^{\tilde{\eta}} d\eta\,\cosh(y-\eta)\,e^{\mu/T}\,
e^{-\tilde{\alpha}\cosh(y-\eta)}\;.
\end{eqnarray}
\section*{The Parametrization of the Chemical Potential}
Spectra of hadrons $k$ containing $q_k$ up or down valence quarks and $s_k$
strange quarks are fully characterized, besides the two flow components, by the
temperature $T=1/\beta$ and the fugacity $\lambda_k = \gamma_{\rm s}^{|s_k|}\:
e^{\beta\mu_{\rm s}s_k}\:e^{\beta\mu_{\rm q}q_k}$ (see the section on relative
chemical equilibrium).

J.~Sollfrank et al.~[2] found in a chemical analysis of S+S collisions, using
rapidity integrated total multiplicities, a vanishing chemical potential
$\mu_{\rm s}\approx0$ of the strange quarks and an almost complete saturation
$\gamma_{\rm s}\approx1$ of the phase space of strange particles.
It is doubtful that these parameters can be established in a purely hadronic
environment, therefore we adopt as a working hypothesis the formation of a QGP
in the early stages of the collision. This leads to $\mu_{\rm s} =0$
independent
of $T$, $\mu_{\rm q}$ and also of $r$ and $z$ (resp.~$\eta$). We further assume
that these values are maintained during the process of hadronisation. (For a
detailed discussion of this assumption see the last reference [1].) Our
objective is to completely fix the baryon chemical potential by using only the
hyperon spectra, and then test whether other measured hadron distributions can
be reproduced in shape as well as normalization.

The width of the rapidity distributions is mainly determined by the
longitudinal
flow component. It can be extracted with very little sensitivity to $T$ and
transverse flow from the shape of the pion rapidity spectrum [5]. In this way
we
obtain a maximum fluid rapidity (see (\ref{endresult})) of $\tilde{\eta}=1.75$.
In a first attempt we neglect the transverse flow ($\rho=0$) and choose
$T=200\pm20\,{\rm MeV}$, consistent with the slope of transverse mass
distributions. We are aware of the fact that this value is too high to be
interpreted as the real freeze-out temperature directly, but in a hadron gas a
temperature of this order seems to be necessary to ensure strangeness
neutrality
[1,2,9].

\begin{figure}
\centerline{
\parbox[t]{7cm}{
\epsfxsize=7truecm
\epsfbox{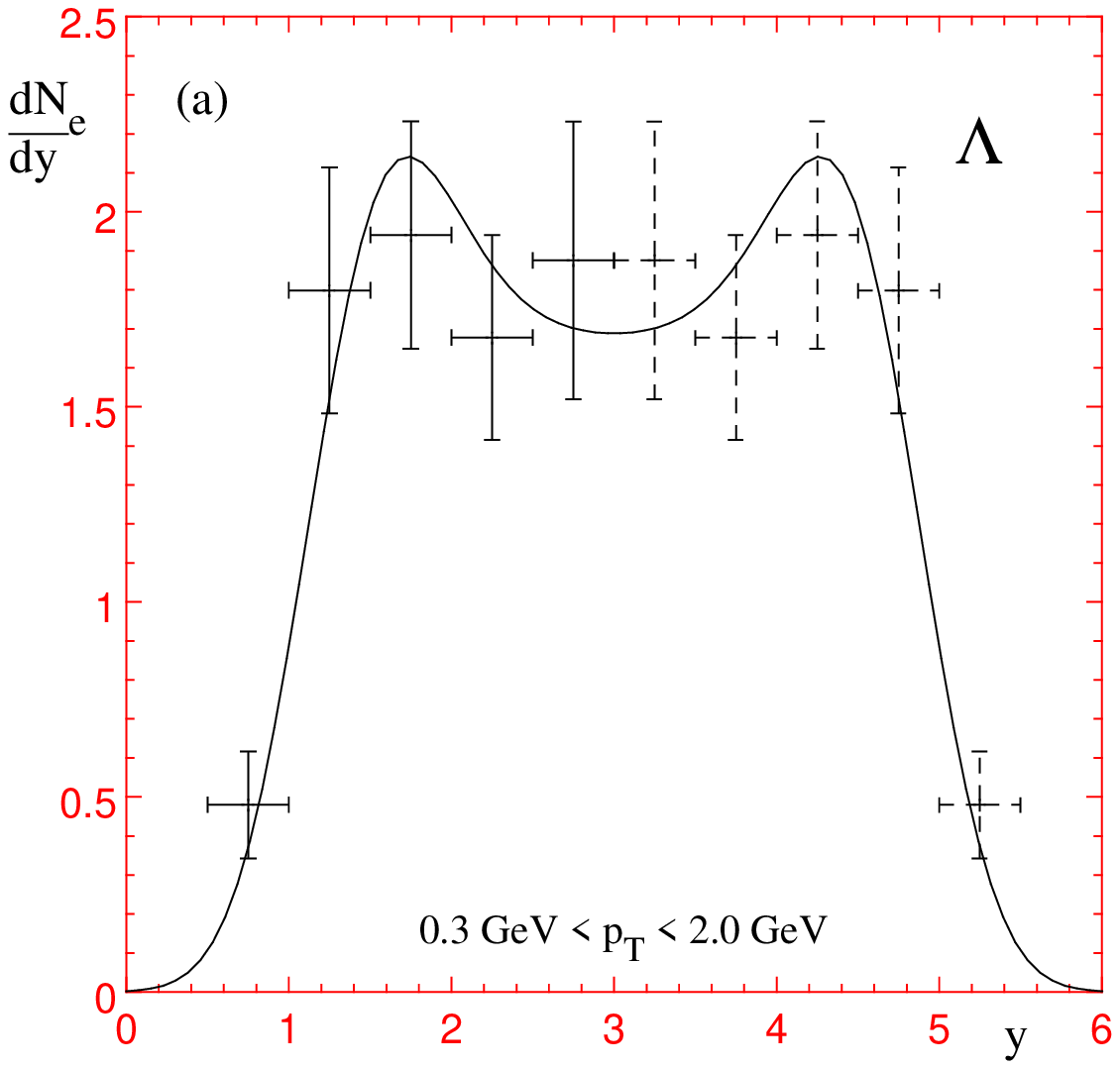} }  \parbox[t]{4mm}{\hfill}
\parbox[t]{7cm}{
\epsfxsize=7truecm
\epsfbox{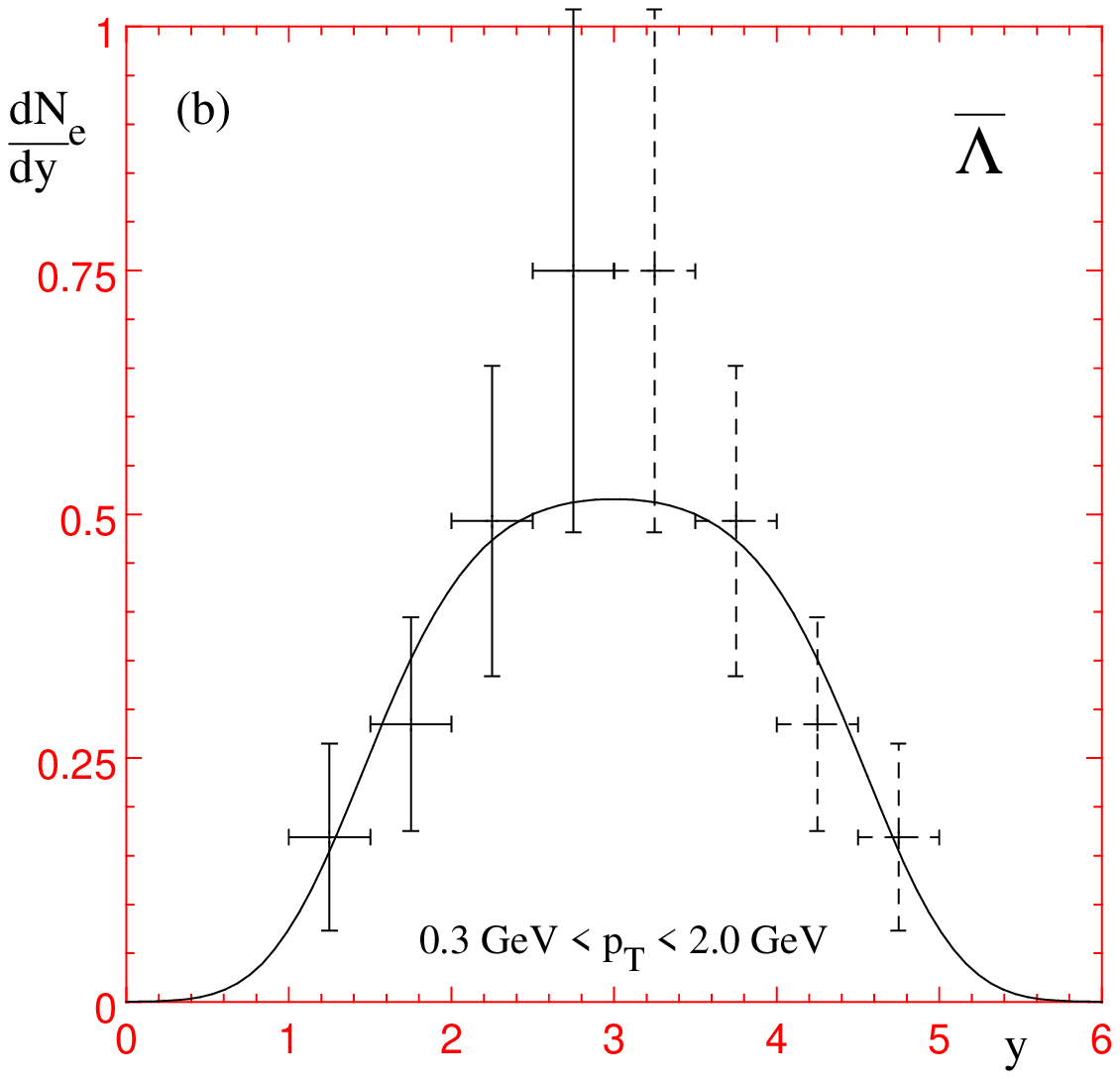} } }
\renewcommand{\baselinestretch}{0.9}
\caption{\label{hyperrap}
Rapidity distribution of $\Lambda$- and $\bar{\Lambda}$-hyperons.
The theoretical curve with our quartic parametrized chemical potential is
fitted to data of NA35 S+S 200$\,$A$\,$GeV [4].}
\vspace*{-2mm}
\end{figure}

Both protons and $\Lambda$-hyperons possess a quite broad rapidity distribution
and peak in the fragmentation regions. This suggests a concave parametrization
for $\mu_{\rm q}(\eta)$. We checked several possibilities and found that a
quartic ansatz of the form
\be \label{quartic}
\mu_k(\eta) = (\mu_{\rm q}^0 + A\,\eta^4)\,q_k\;,\quad \mu_{\rm q}^0=  59\pm11
\,{\rm MeV}\;,\quad A=11.5\pm0.5\,{\rm MeV}
\ee
together with $\tilde{\eta} = 1.75$ is able to describe both $\Lambda$- and
$\bar{\Lambda}$-distributions simultaneously (Fig.~\ref{hyperrap}~a-b).
\renewcommand{\baselinestretch}{1.0}
The central data points of the $\bar{\Lambda}$'s are barely reached only, but
they have the largest error bars. We used the fact that both spectra are
measured in the same $p_\bot$-window, whereby the integrals over the boostangle
$\eta$ cancel in the ratio of the distributions at midrapidity. From the
central
normalization of the $\Lambda$-distribution we also determine
$\tau_{\rm f} R_{\rm f}^2 = 20\pm4\:{\rm fm}^{3}$.
The contribution from $\Sigma^0$ decays to the $\Lambda$ spectrum is taken into
account in the theory by an additional degeneracy factor of 2.
\section*{The Results}
\begin{figure}
\centerline{
\parbox[t]{7cm}{
\epsfxsize=7truecm
\epsfbox{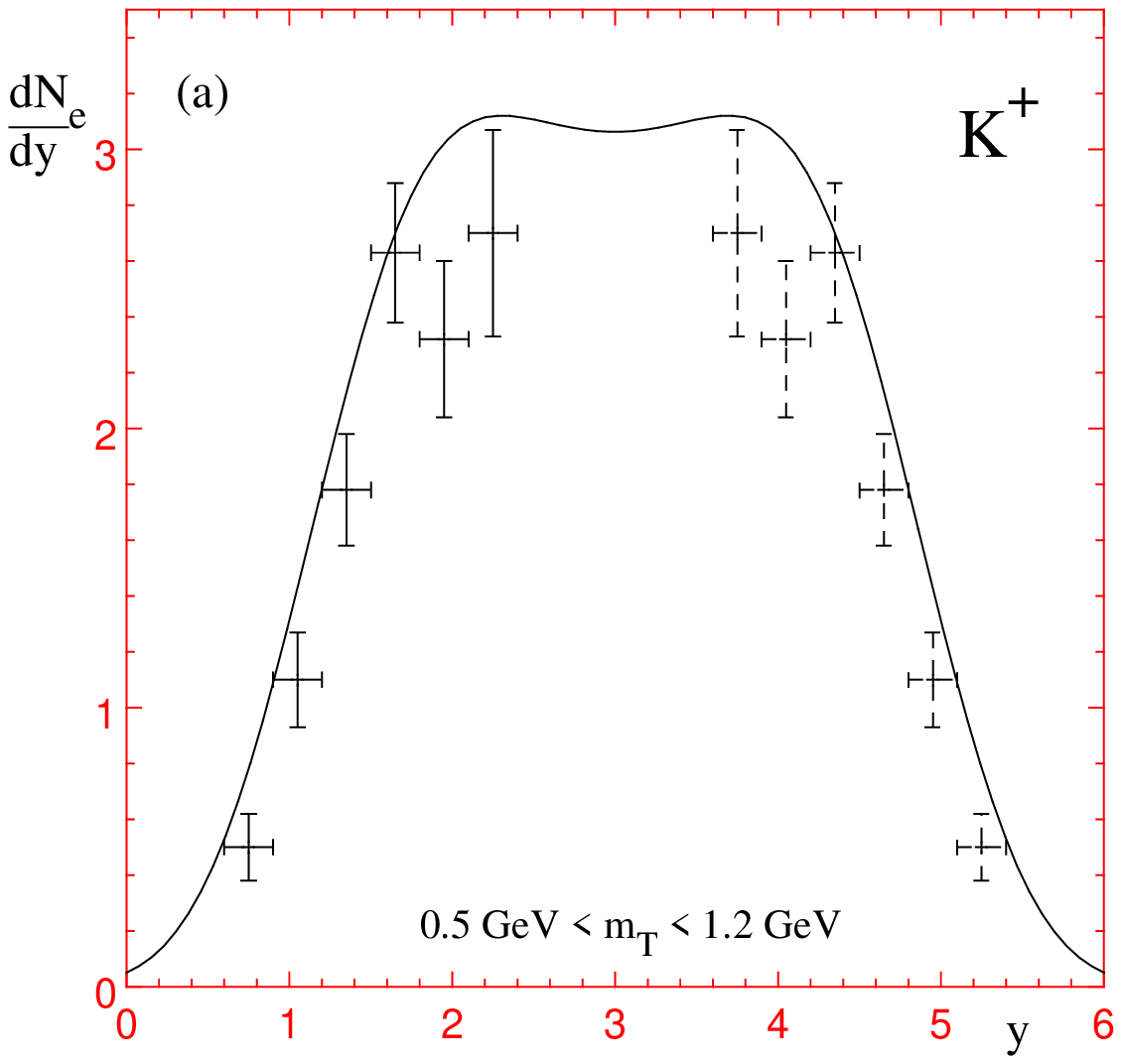} } \parbox[t]{4mm}{\hfill}
\parbox[t]{7cm}{
\epsfxsize=7truecm
\epsfbox{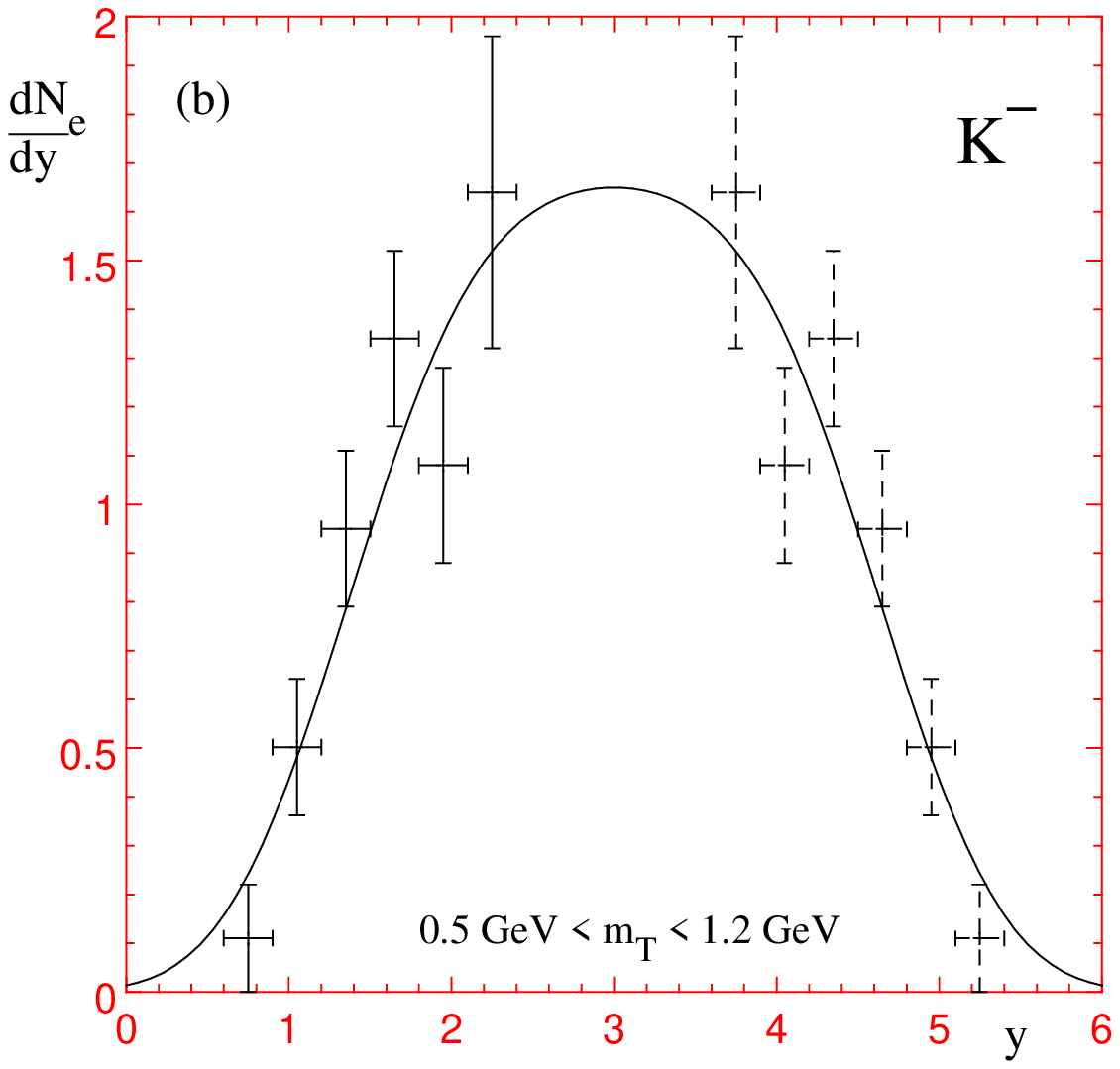} } }
\centerline{
\parbox[t]{7cm}{
\epsfxsize=7truecm
\epsfbox{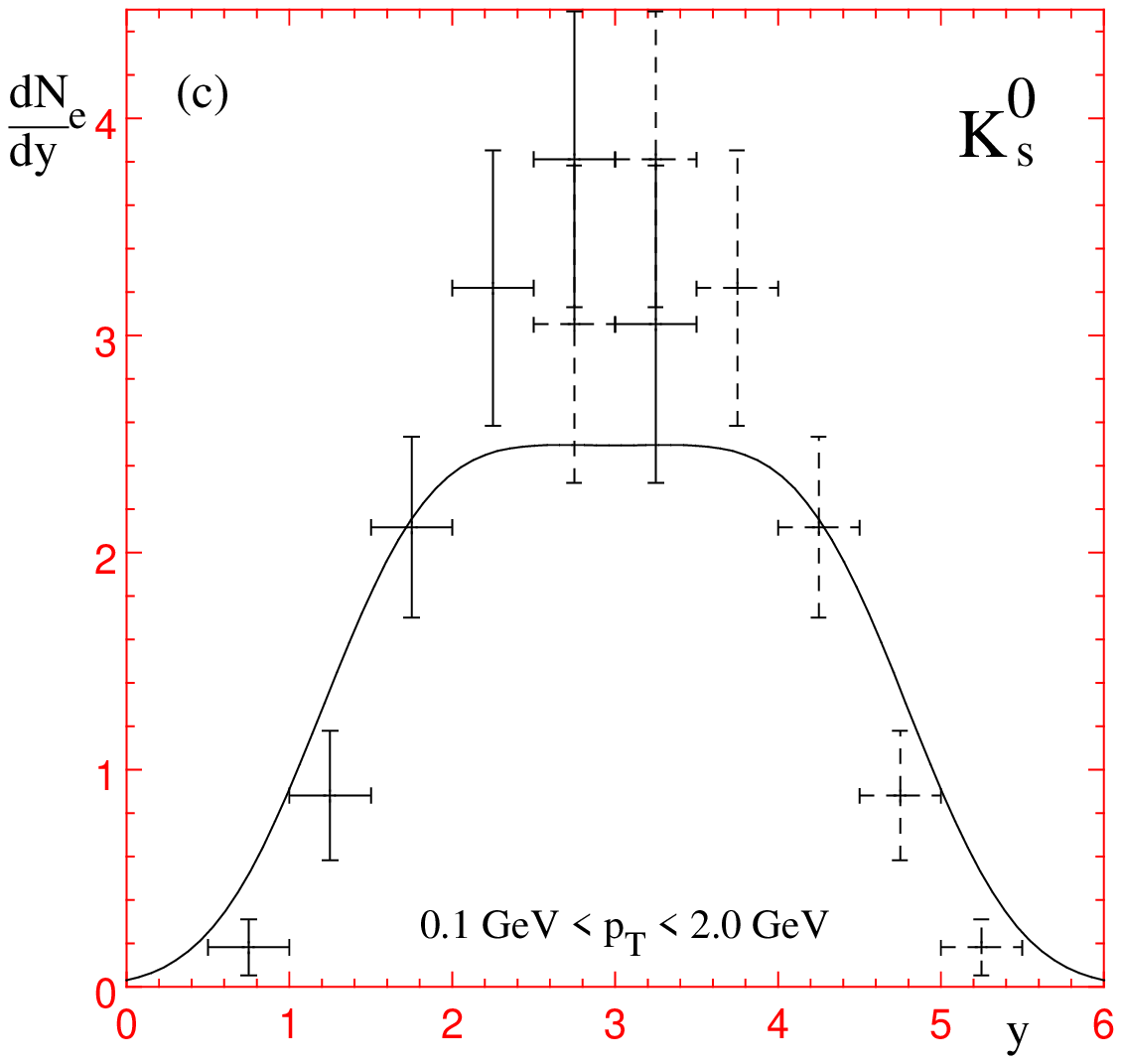} } \parbox[t]{4mm}{\hfill}
\parbox[t]{7cm}{
\epsfxsize=7truecm
\epsfbox{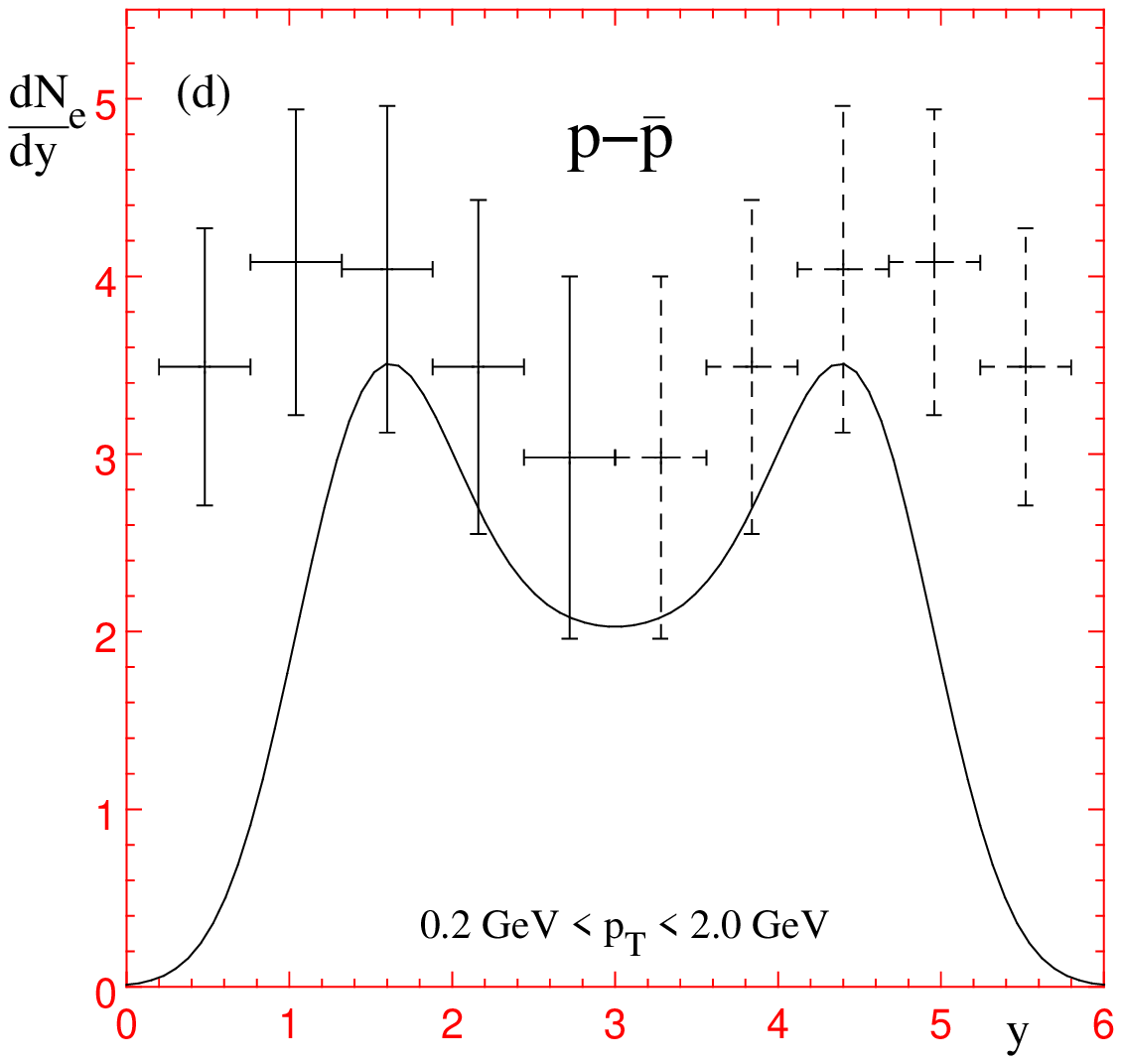} } }
\centerline {
\parbox[t]{7cm}{
\epsfxsize=7truecm
\epsfbox{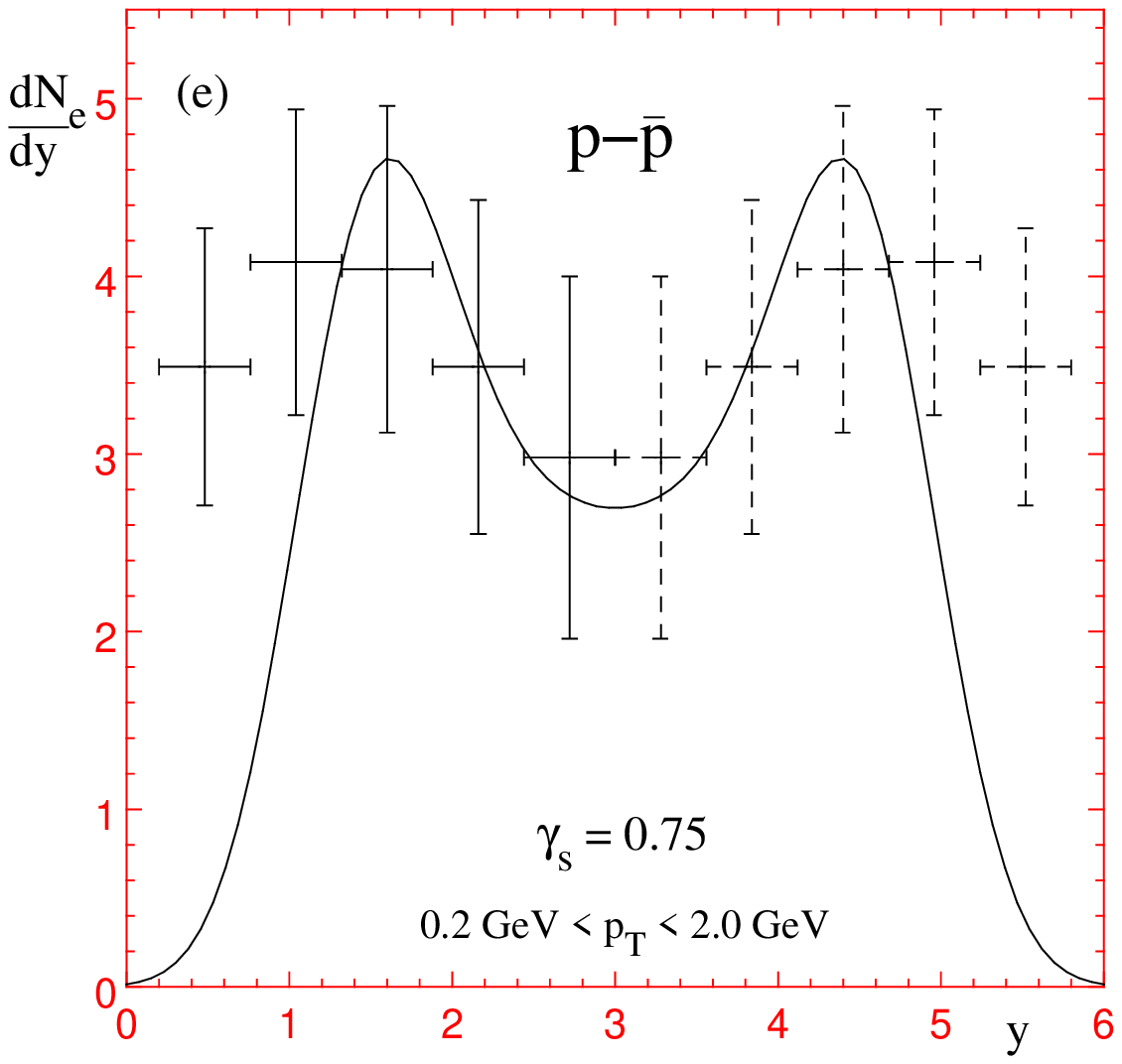} }  \parbox[t]{4mm}{\hfill}
\raisebox{6.5cm}{\begin{minipage}[t]{7truecm}
\hspace*{7mm}
\begin{minipage}[t]{5.8cm}
\renewcommand{\baselinestretch}{0.9}
\caption {\label{resrap}
Rapidity distribution of kaons and protons in comparison with data of NA35 for
S+S at 200$\,$A$\,$MeV [4].
The absolute normalizations are {\it not} fitted, but calculated using
$T=200\,{\rm MeV}$, $\tilde{\eta}=1.75$ and the quartic parametrization
for the chemical potential. Figure e) represents the proton distribution
assuming an undersaturation of overall strangeness $\gamma_{\rm s}=0.75$.}
\renewcommand{\baselinestretch}{1.0}
\end{minipage}
\end{minipage} } }
\end{figure}

Integrating $\mu(\eta)$ over the fluid (or space-time) rapidity $\eta$, we
determine an average baryo\-chemical potential of
$\bar{\mu}_{\rm B} = 3\bar{\mu}_{\rm q} = (240\pm40)\:{\rm MeV}$.
This is in good agreement with the findings of J.~Sollfrank et al.~[2],
$\mu_{\rm B} = (264\pm72)\:{\rm MeV}$ who used a chemical analysis of the
$4\pi$-integrated multiplicities. But the most important result is that we are
able to also reproduce all other available hadron spectra
(Fig.~\ref{resrap}~a-d). Not only the shape of the distributions coincides with
the experiment, but also (except for the pions which will be discussed
separately below) the normalization is reproduced correctly without
adjustments.
(Due to error propagation some of our kaon curves may, however, have up to 35\%
uncertainty in normalization.) This is a dramatic improvement over the results
obtained in Ref.~[5] with rapidity independent chemical potentials.

The calculated proton distribution (Fig.~\ref{resrap}~d) is clearly narrower
than the experimental result, moreover the normalization is a little too low.
The first feature is expected because protons are unlikely to be completely
thermalized, especially in the extreme kinematic domains, due to nuclear
transparency in particular in the transverse nuclear surface. On the other hand
a slight undersaturation of the overall strangeness by a factor $\gamma_{\rm s}
= 0.75$ fully corrects the normalization problem (Fig.~\ref{resrap}~e). Since
all the other hadrons carry only one $s$- or $\bar{s}$-quark, this leaves the
spectra unchanged, if the factor $\tau_{\rm f}\,{\rm R}_{\rm f}^2$ is increased
correspondingly by a factor $4/3$.

We conclude from this agreement with experiment that a thermodynamic
description
will work well in a rather wide rapidity interval if dependencies of the
thermodynamic parameters on rapidity are taken into consideration. The
interesting observation here is that it is apparently sufficient to consider a
rapidity dependence of the baryon chemical potential, keeping $T$, $\mu_{\rm
s}=0$ and $\gamma_{\rm s}$ fixed. On the other hand this means that one has to
be careful when applying thermodynamic models to data which are not restricted
to a narrow and central kinematic domain, because then this rapidity dependence
is integrated over with different weights for different particle species.
As a further result the strong dependence of strange particle ratios on the
rapidity window which was observed by NA36 in S+Pb collisions [3] can be
qualitatively explained by our model in a very natural way.

\begin{figure}
\centerline {
\parbox[t]{7cm}{
\epsfxsize=7truecm
\epsfbox{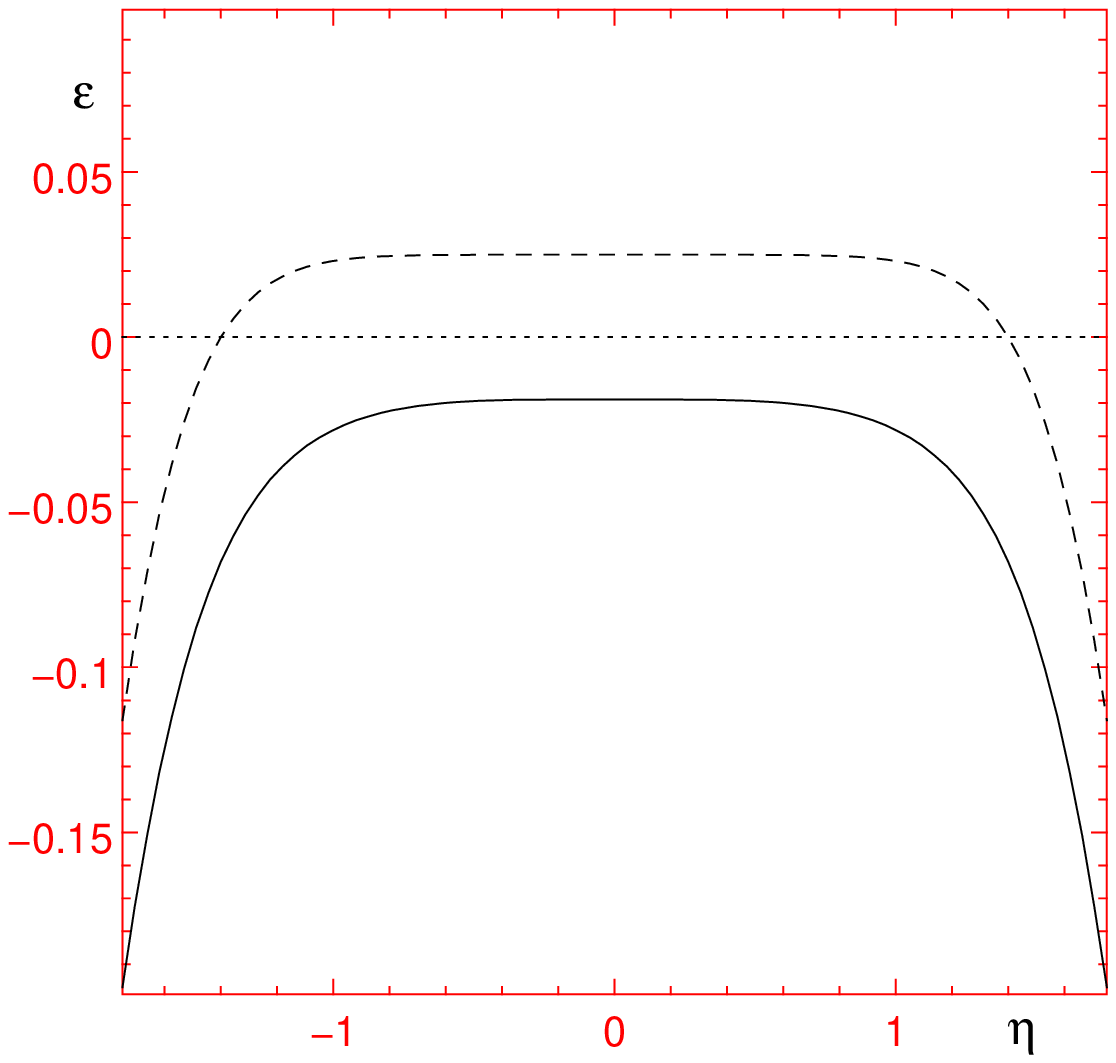} }  \parbox[t]{4mm}{\hfill}
\raisebox{6.5cm}{\begin{minipage}[t]{7truecm}
\hspace*{7mm}
\begin{minipage}[t]{5.8cm}
\renewcommand{\baselinestretch}{0.9}
\caption{\label{strange}
Deviation from local strangeness neutrality. The quantity $\epsilon$ (23)
is plotted as a function of space-time rapidity $\eta$ for $T=200\,{\rm MeV}$,
$\gamma_{\rm s}=1$ and the parametrization (22)
for the chemical potential. The dashed curve is calculated using a mass cut in
the hadronic resonance spectrum at $m_{\rm cut}=1.8\;{\rm GeV}$, the solid one
represents $m_{\rm cut}=2.0\;{\rm GeV}$.}
\end{minipage}\end{minipage} } }
\vspace*{-10mm}
\end{figure}
\renewcommand{\baselinestretch}{1.0}
To test the important condition of strangeness neutrality we define
\be \label{epsilon}
\epsilon(\eta) = \frac{\la\bar{s}\ra(\eta) - \la s\ra(\eta)}{\la s\ra(\eta)}
\ee
as a measure for the local deviation. Fig.~\ref{strange} shows the result
plotted in the relevant range of $\eta$ for two different cuts in the hadron
mass spectrum. We recognize that strangeness neutrality can not be guaranteed
locally, though $\epsilon(\eta)$ is flat and almost zero in
$-1.2\le\eta\le1.2$.
On the other hand the clear experimental differences between the rapidity
densities of different hadrons can be seen as an indication that a violation
might indeed be realized locally. In a global sense, i.e.~integrated over
rapidity $\eta$, we obtain

\vspace{3mm}
\centerline{
\begin{tabular}{|c|c|c|c|c|}\hline
$\gamma_{\rm s}$&\multicolumn{2}{|c|}{1.0}&\multicolumn{2}{|c|}{0.75}\\ \hline
$m_{\rm cut}$ [GeV] & 1.8 & 2.0 & 1.8 & 2.0 \\ \hline
$\epsilon_{\rm global}$ & 0.03 & -0.16 & 0.09 & -0.09 \\ \hline
\end{tabular}$\;\;$.}\vspace{2mm}
If we take into consideration the uncertainties in temperature, chemical
potential and a possible asymmetry in the data tables between meson and baryon
resonances of high rest masses (due to different efficiencies for their
identification), we conclude that we have no problem in ensuring at least
global strangeness neutrality.
\section*{Transverse flow}
Despite this very good and consistent agreement of our model with experiment
this comparison has one substantial shortcoming: the assumed temperature is
really too high for consistency of the model. The spatial overlap of hadrons at
$T=200\,{\rm MeV}$ is quite large already, and indeed lattice QCD simulations
now firmly predict a QGP phase transition already at $T=150\,{\rm MeV}$ [10].
The assumption of a system of non-interacting particles freezing out at
$T=200\,{\rm MeV}$ is thus more than dubious. These were the decisive reasons
for the authors of [2,5] to conclude that the flat slope of the observed
$m_\bot$-spectra is partially caused by a blueshift due to collective
transverse
flow. Already $\rho\simeq0.3$ reduces the true temperature from
$T\simeq200\,{\rm MeV}$ to $T\simeq150\,{\rm MeV}$, and we will use this
average
value in the following considerations.

\begin{figure}
\centerline{
\parbox[t]{7cm}{
\epsfxsize=7truecm
\epsfbox{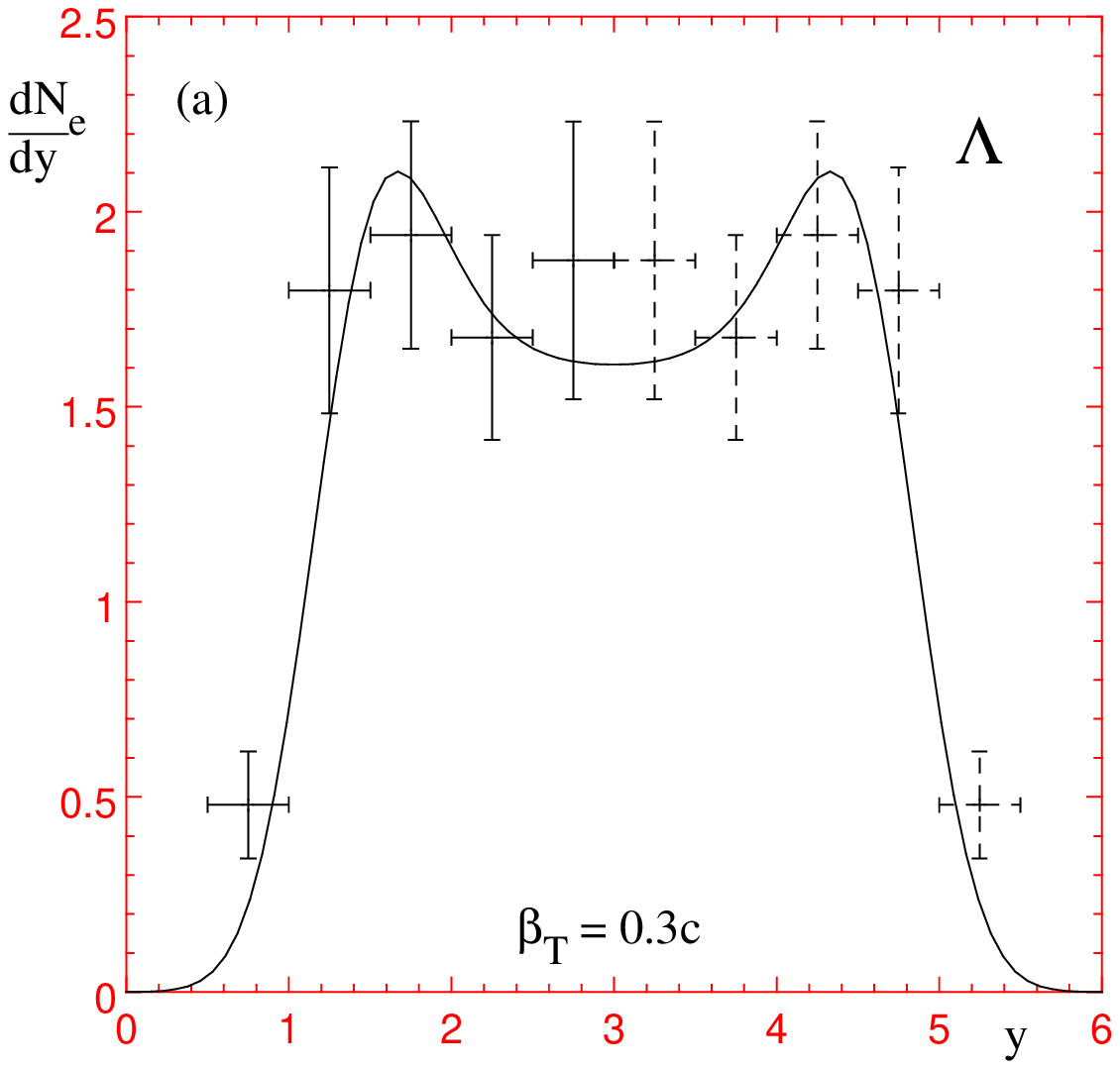} } \parbox[t]{4mm}{\hfill}
\parbox[t]{7cm}{
\epsfxsize=7truecm
\epsfbox{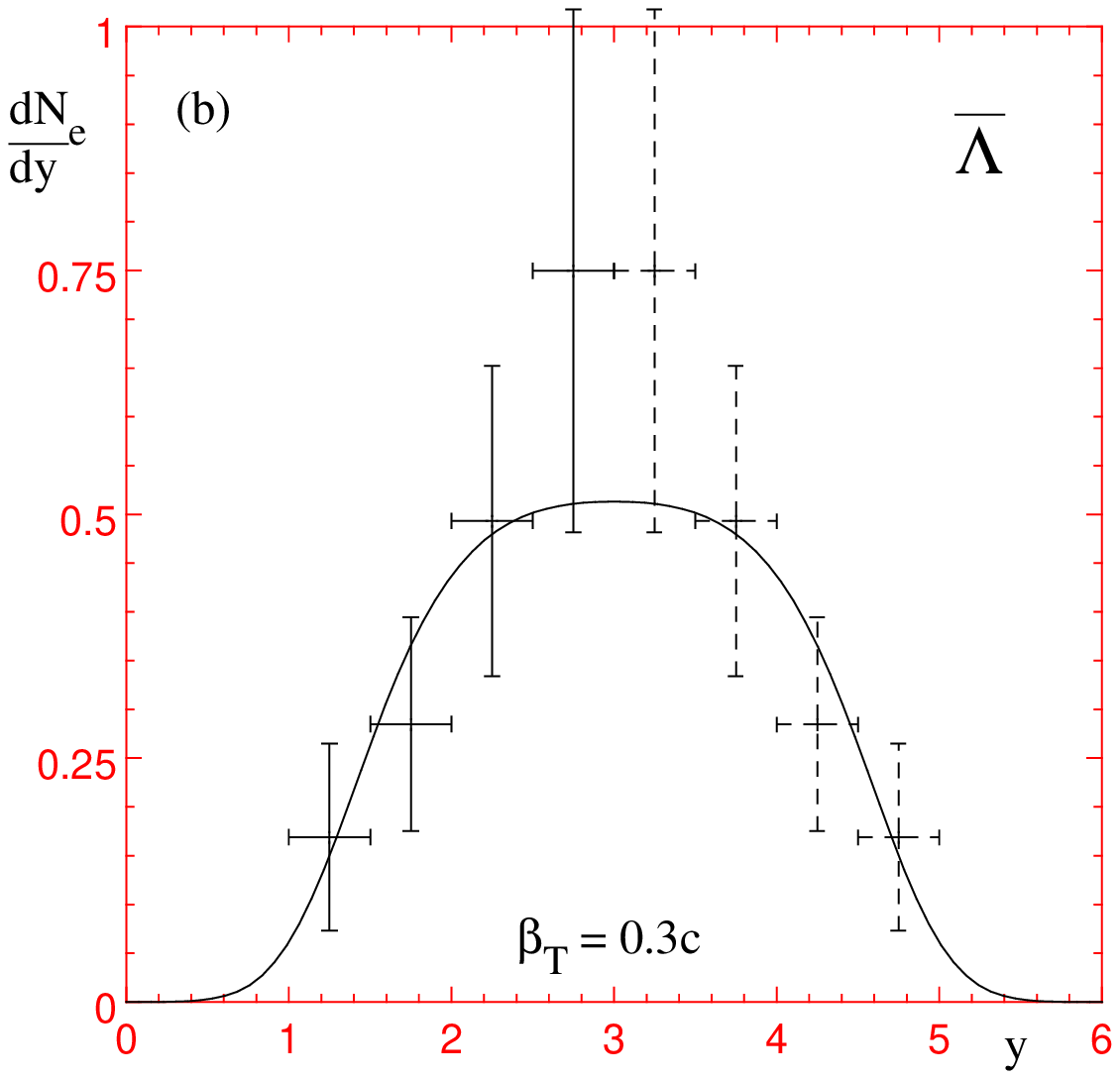} } }
\centerline{
\parbox[t]{7cm}{
\epsfxsize=7truecm
\epsfbox{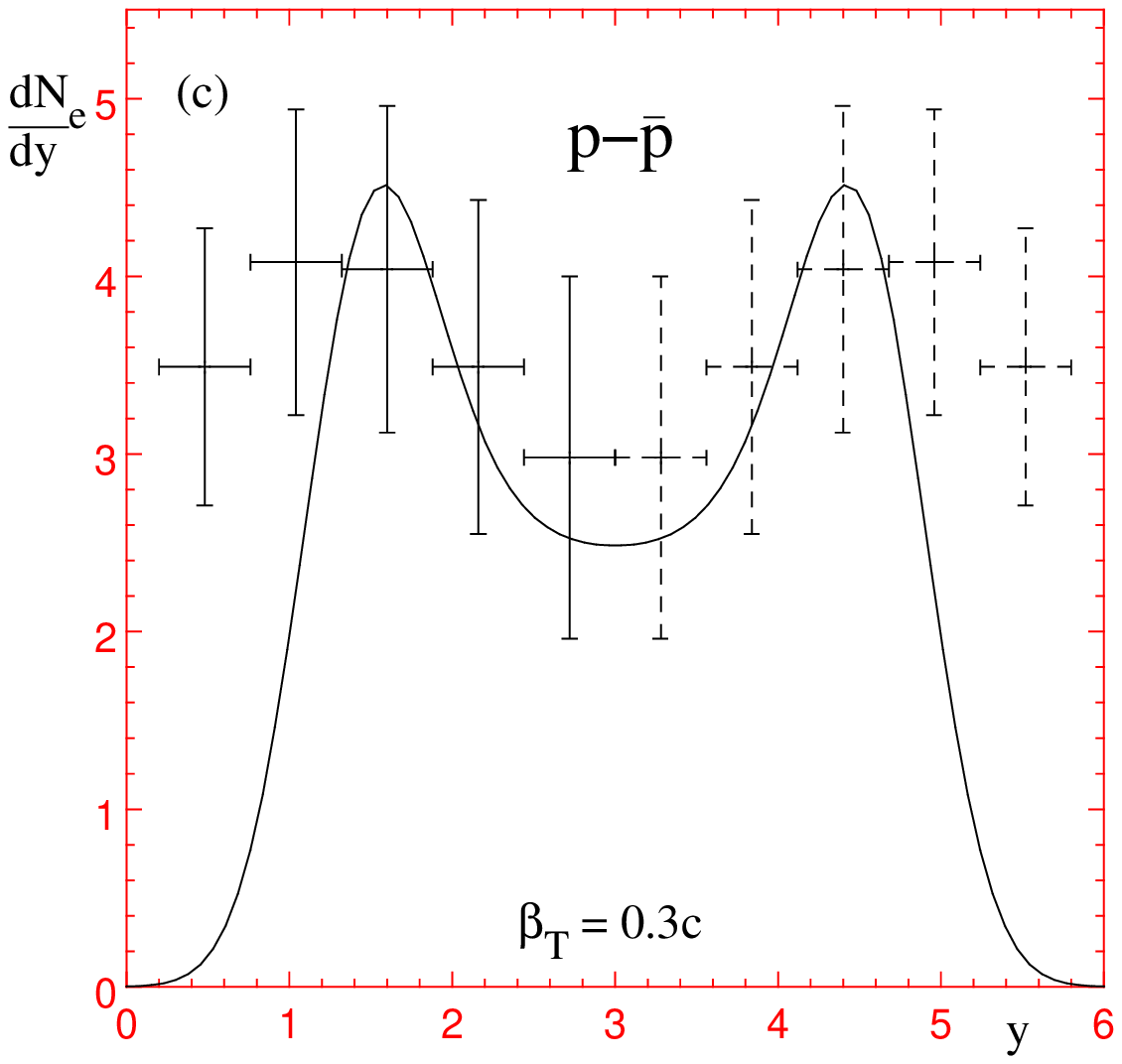} } \parbox[t]{4mm}{\hfill}
\parbox[t]{7cm}{
\epsfxsize=7truecm
\epsfbox{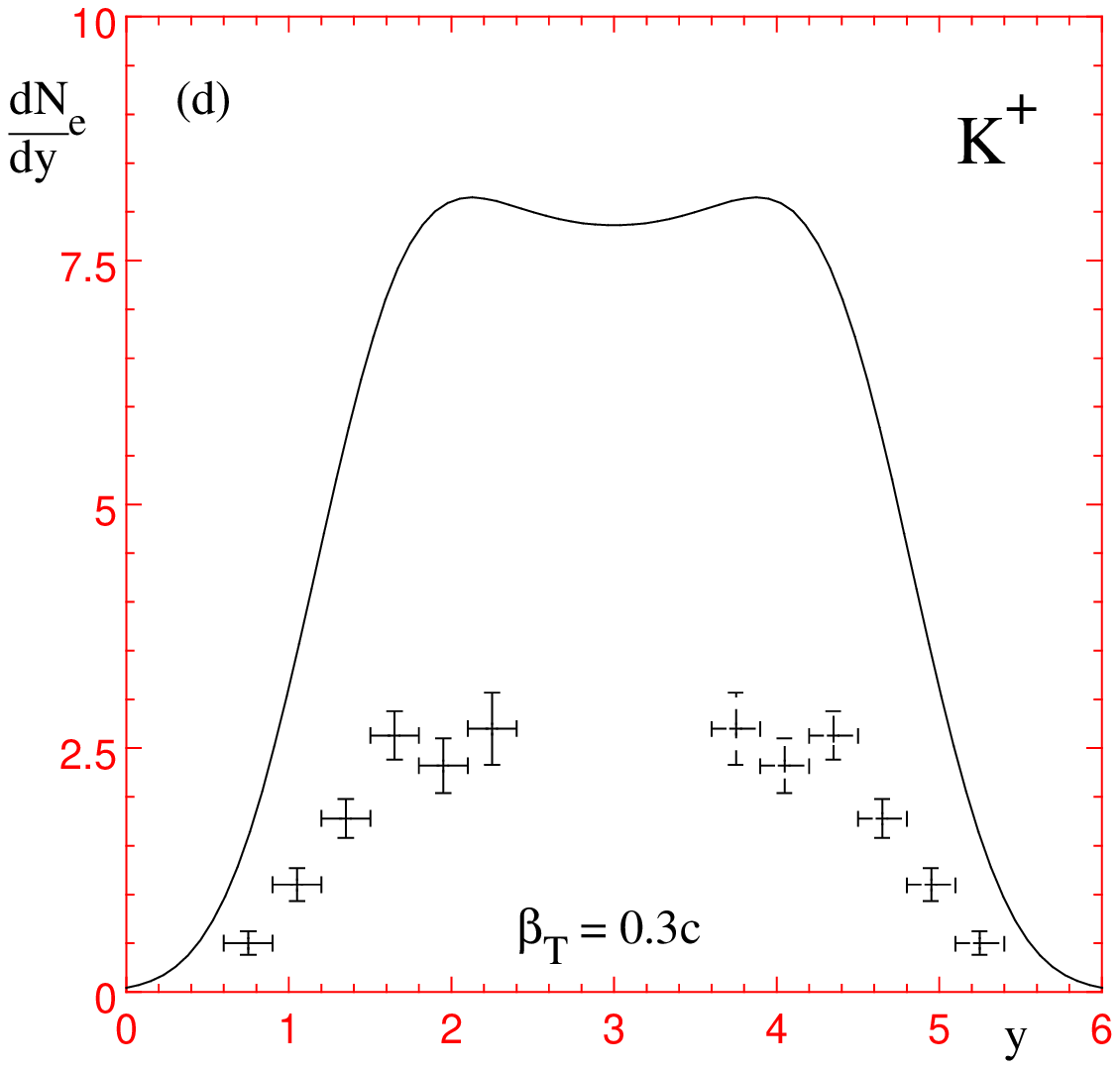} } }
\centerline {
\parbox[t]{7cm}{
\epsfxsize=7truecm
\epsfbox{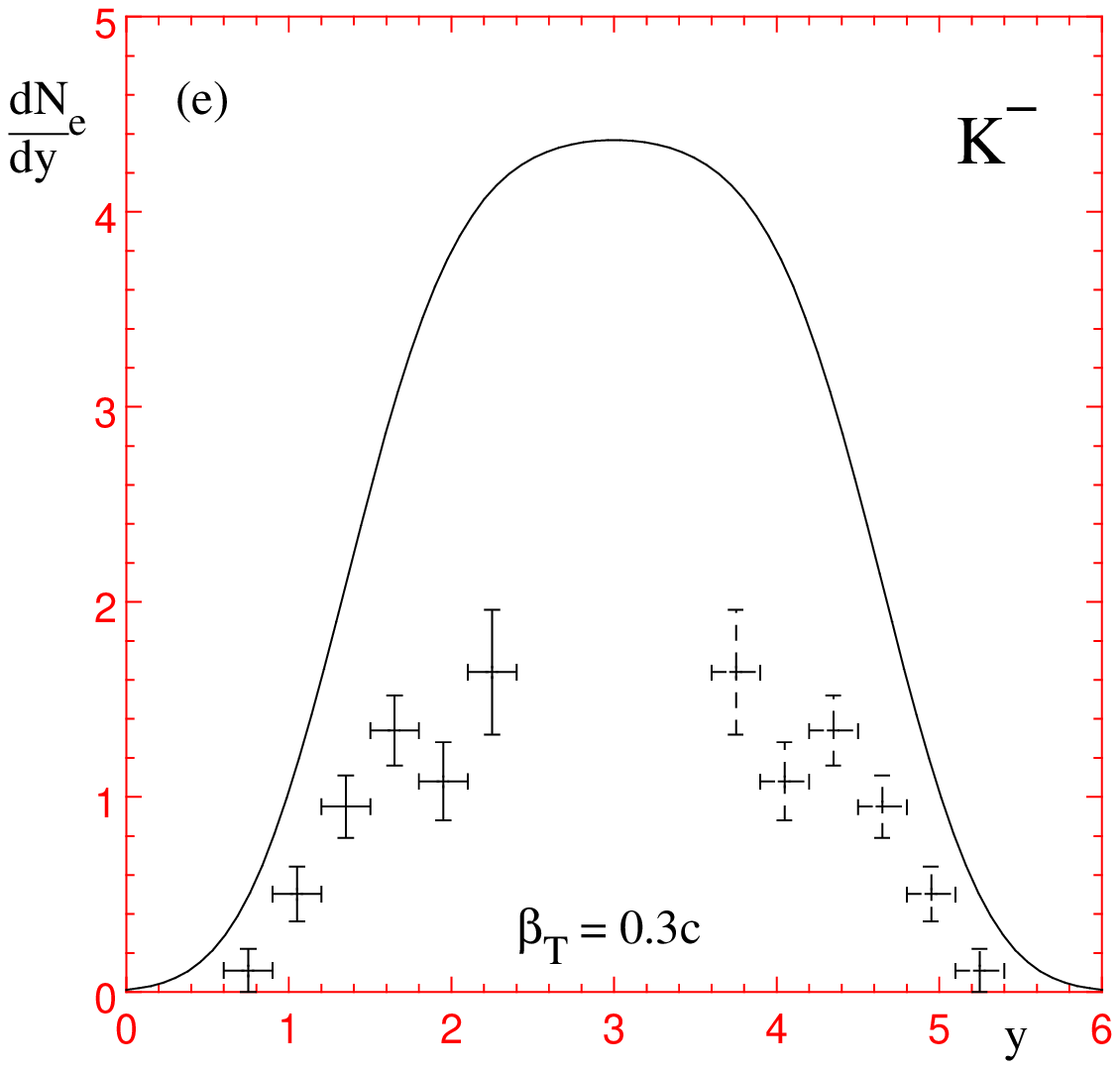} } \parbox[t]{4mm}{\hfill}
\raisebox{6.5cm}{\begin{minipage}[t]{7truecm}
\hspace*{7mm}
\begin{minipage}[t]{5.8cm}
\renewcommand{\baselinestretch}{0.9}
\caption {\label{flowrap}
Rapidity distributions of kaons and protons in comparison with data of NA35 S+S
at 200$\,$A$\,$MeV [4].
The absolute normalizations in the cases c)-e) are {\it not} fitted, but
calculated using $T=150\,{\rm MeV}$, $\tilde{\eta}=1.75$ and the quartic
parametrization for the chemical potential. Please note that
$\gamma_{\rm s}=\gamma_{\rm m}=1$ was assumed.}
\renewcommand{\baselinestretch}{1.0}
\end{minipage}\end{minipage} } }
\end{figure}
We again determine $\mu_{\rm q}^0$ and $A$ as described above from the hyperon
spectra, now leading to $A=8.0\pm0.5\,{\rm MeV}$ and $\mu_{\rm q}^0 \simeq
43\,{\rm MeV}$. The lower temperature reduces the thermal smearing whereby all
the theoretical rapidity spectra become somewhat narrower. The calculated
proton
distribution now fits the experimental data (except for the disagreement near
the projectile and target rapidities discussed already above), even with
$\gamma_{\rm s}=1$ (Fig.~\ref{flowrap}~c).
\begin{figure}
\centerline{
\parbox[t]{7cm}{
\epsfxsize=7truecm
\epsfbox{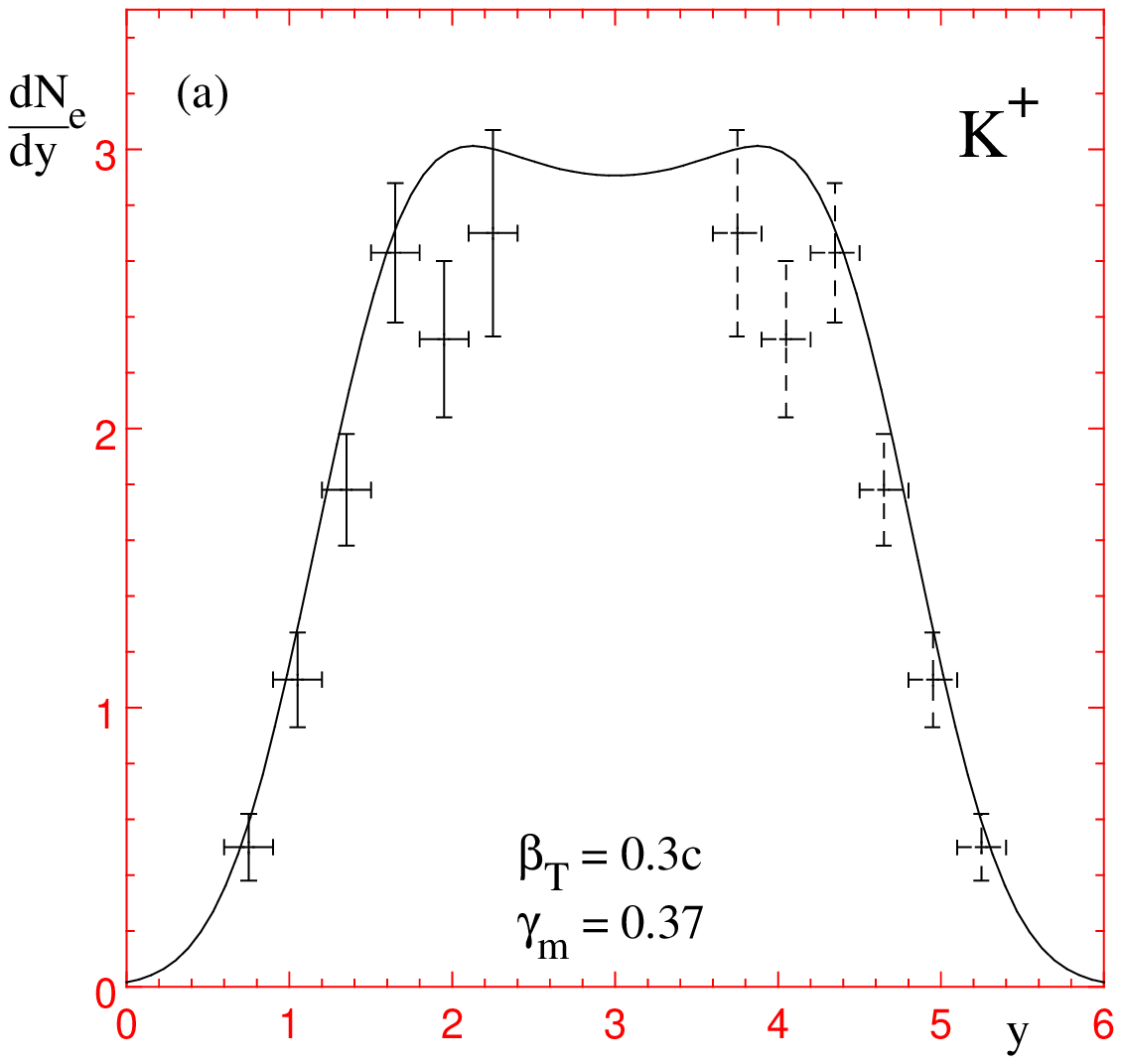} }  \parbox[t]{4mm}{\hfill}
\parbox[t]{7cm}{
\epsfxsize=7truecm
\epsfbox{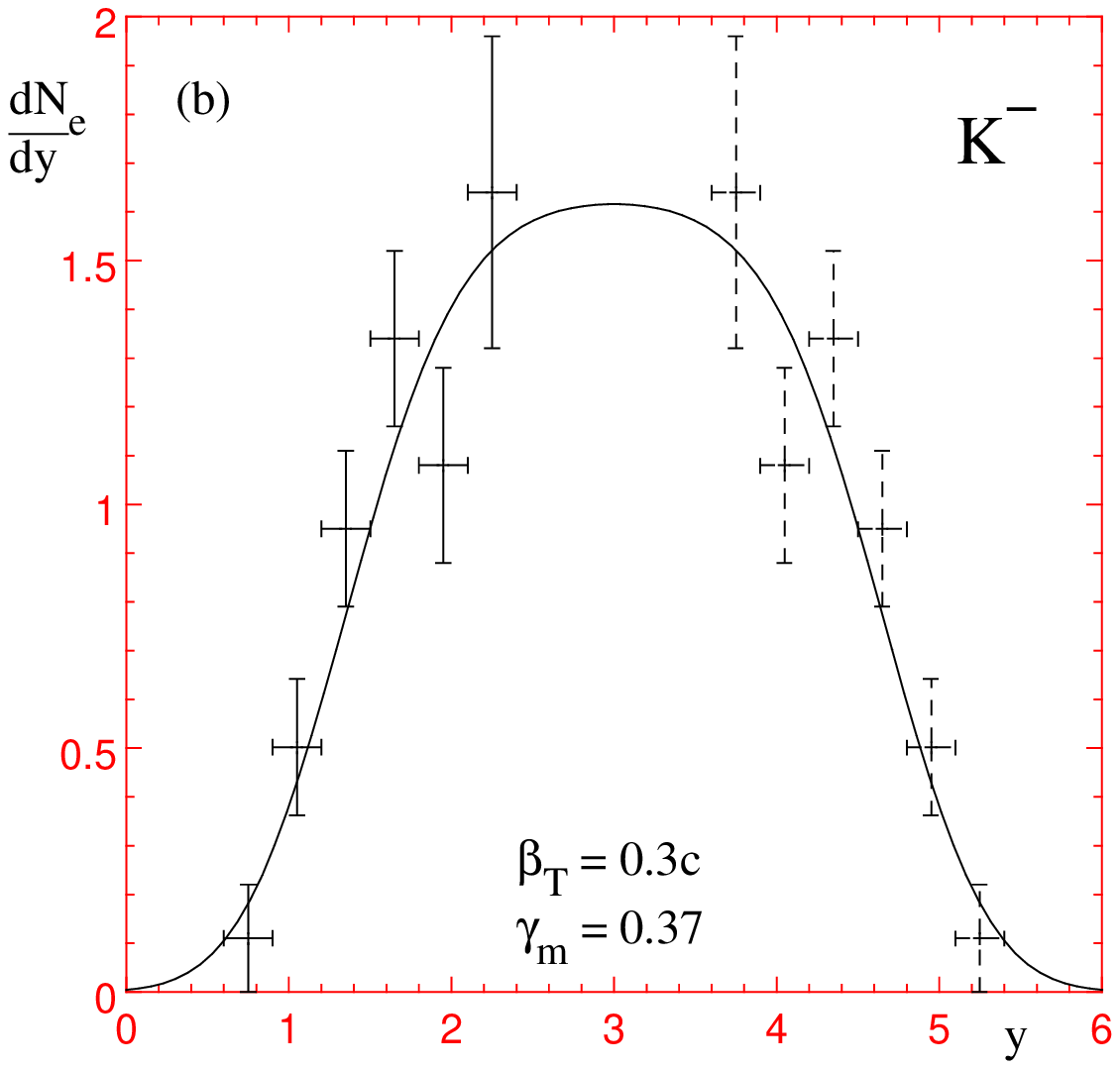} } }
\renewcommand{\baselinestretch}{0.9}
\caption{\label{kaonplot}
Rapidity distribution of kaons in comparison with data of NA35 S+S
200$\,$A$\,$GeV [4]. The theoretical curve is calculated using $T=150\,{\rm
MeV}$, $\tilde{\eta}=1.75$ and $\gamma_{\rm m}=0.37$.}
\vspace*{-2mm}
\end{figure}
\renewcommand{\baselinestretch}{1.0}

With kaons, however, we now have a serious problem: Although the shape is
correct for all three charge states, the theoretical normalization exceeds the
data by a common factor of approximately 2.7 (Fig.~\ref{flowrap}~d-e). As a
consequence the theoretical spectra also strongly violate overall strangeness
neutrality.
However, there is a very simple remedy for this. Following the ideas first laid
out in Ref.~[1d] and worked out in Ref.~[1c], one can assume a breaking of the
relative chemical equilibrium between the mesons and baryons by introducing a
meson/baryon suppression factor $\gamma_{\rm m} <1$. It is quite a surprise
that
by choosing $\gamma_{\rm m} =0.37$ we can simultaneously correct the
normalization of all kaon spectra (Fig.~\ref{kaonplot}) {\it and} recover
global
strangeness neutrality:

\vspace{3mm}\centerline{
\begin{tabular}{|c|c|c|} \hline
    $ m_{\rm cut}$ [GeV] & 1.8 & 2.0 \\ \hline
    $\epsilon_{\rm global}$ & -0.05 & -0.13 \\ \hline
\end{tabular}$\;.$}\vspace{2mm}
\section*{Excess Entropy}
\begin{figure}
\centerline{
\parbox[t]{7cm}{
\epsfxsize=7truecm
\epsfbox{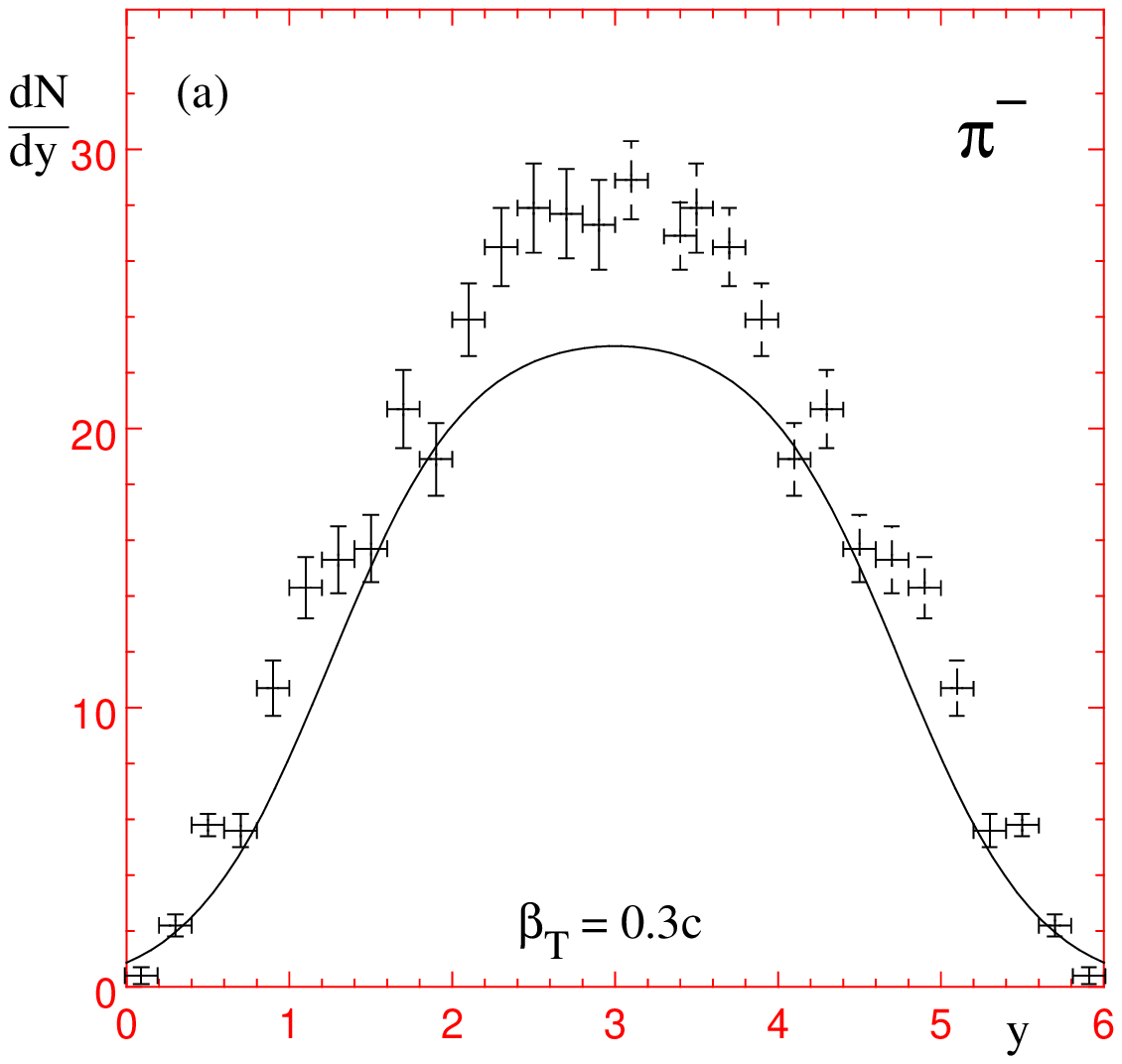} } \parbox[t]{4mm}{\hfill}
\parbox[t]{7cm}{
\epsfxsize=7truecm
\epsfbox{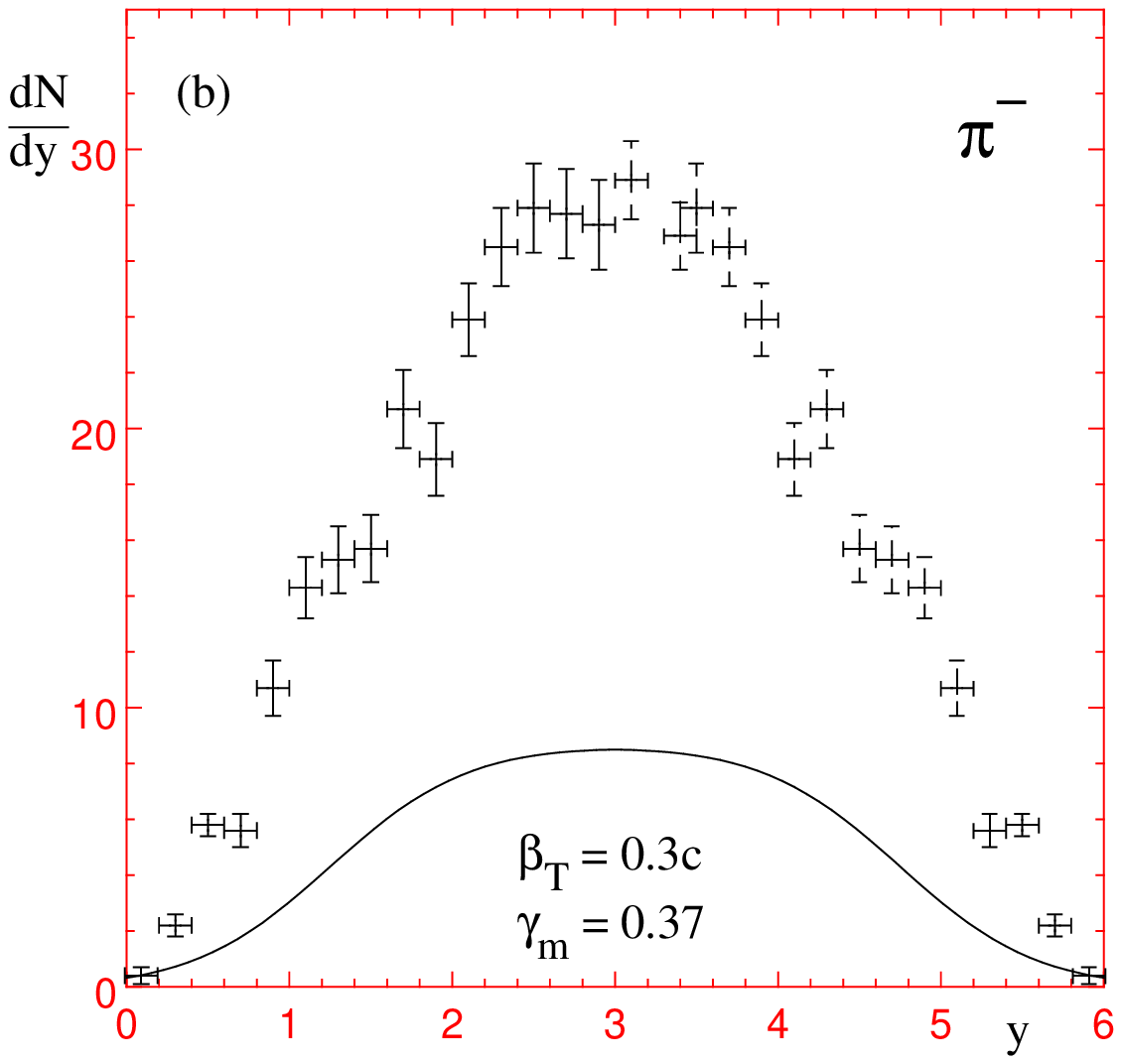} } }
\renewcommand{\baselinestretch}{0.9}
\caption{\label{pions}
Rapidity distribution of pions. The theoretical curve with our quartic
parametrized chemical potential and a transverse flow of $\beta_{\rm T} =
\tanh\rho = 0.3$ is compared to data of NA35 S+S 200$\,$A$\,$GeV [4]. Figure b)
is based on a meson baryon suppression factor $\gamma_{\rm m}= 0.37$.}
\vspace*{-2mm}
\end{figure}

In both analyses (with and without transverse flow) we used the pion rapidity
spectra to extract the maximum  flow rapidity, but did not check their
normalization. It turns out that in both cases we seriously underpredict the
pion multiplicity (typically by a factor 2-3). While in the case with
transverse
flow and $\gamma_{\rm s}=\gamma_{\rm m}=1$ we are not too much off (see
fig.~\ref{pions}~a), after introducing the meson suppression factor
$\gamma_{\rm
m}=0.37$ (thereby recovering the correct kaon normalization and strangeness
neutrality) we again obtain by a factor 2-3 too few pions. This feature in the
data (too many pions compared to a hadron gas in relative chemical equilibrium)
was observed before [1,2,11] and is confirmed by the completely different
experimental analysis reported by M.~Ga\'zdzicki at ``Strangeness '95''
[12]. It is interpreted as an indication for excess entropy in the data which
might originate from the gluonic entropy in the QGP [1]. Why this extra entropy
shows up mainly in the form of pions is not yet clear. Their nature as nearly
massless chiral goldstone bosons may play a role here, but no dynamical
explanation has been given so far.

\renewcommand{\baselinestretch}{1.0}
\section*{Summary}
We showed that a rather simple thermodynamic model with a suitably parametrized
rapidity dependent baryochemical potential can reproduce the measured
distributions of strange particles resulting from S+S collisions at
200$\,$A$\,$GeV. Consistency was achieved with the shape as well as with the
normalization of the rapidity spectra. Our phenomenologically motivated ansatz
was not able to guarantee strangeness neutrality locally, but globally
strangeness was conserved.
One further result of our model was a qualitative explanation of the strong
dependence of strange particle ratios on the chosen kinematic window.

Resonance decays have not been taken into consideration so far, but will be
considered in the future. They are expected to mostly affect the pion spectra,
but also kaons could be influenced in a non-negligible way. We have
conscisously
kept $\mu_{\rm s}=0$ and $\gamma_{\rm s}$, $\gamma_{\rm m}$ as
rapidity-independent, because otherwise a unique determination of such a large
set of free parameter functions would be impossible to fix from the available
data.

\end{document}